\documentclass{article}

\usepackage[english]{babel}

\usepackage[a4paper,top=2cm,bottom=2cm,left=3cm,right=3cm,marginparwidth=1.75cm]{geometry}

\usepackage[numbers]{natbib}
\usepackage{graphicx}
\usepackage{subcaption}

\usepackage{amsmath}
\usepackage{amsfonts}
\usepackage{amssymb}
\usepackage{amsthm}
\usepackage[textsize=small]{todonotes}
\usepackage{comment}
\usepackage{tabularx}
\usepackage{url}
\usepackage{hyperref}

\usepackage{tikz}
\usetikzlibrary{shapes,arrows,positioning,calc}

\usepackage{environ,tcolorbox}
\NewEnviron{question}{
	\begin{tcolorbox}[colback = yellow]
 \textbf{Comments / Questions:}
 
		\BODY
	\end{tcolorbox}
}

\usepackage{environ,tcolorbox}
\NewEnviron{bullets}{
	\begin{tcolorbox}
 \textbf{Outline / key points:}
 
		\BODY
	\end{tcolorbox}
}

\usepackage{authblk}

\title{Value of risk-contact data from digital contact monitoring apps in infectious disease modeling}
\author[a]{Martijn H. H. {Schoot Uiterkamp}\thanks{Corresponding author. E-mail:  \url{m.h.h.schootuiterkamp@tilburguniversity.edu}.}}
\author[b,c]{Willian J. van Dijk}
\author[d]{Hans Heesterbeek}
\author[e]{Remco~van~der~Hofstad}
\author[b]{Jessica C. {Kiefte-de Jong}}
\author[e]{Nelly Litvak}
\affil[a]{Department of Econometrics and Operations Research, Tilburg University, The Netherlands}
\affil[b]{Department of Public Health and Primary Care, Health Campus The Hague and Leiden University Medical Center, The Netherlands}
\affil[c]{Department of Clinical Epidemiology, Leiden University Medical Center, The Netherlands}
\affil[d]{Department of Population Health Sciences, Utrecht University, The Netherlands}
\affil[e]{Department of Mathematics and Computer Science, Eindhoven University of Technology, The Netherlands}

\begin{document}

\maketitle

\begin{abstract}
In this paper, we present a simple method to integrate risk-contact data, obtained via digital contact monitoring (DCM) apps, in conventional compartmental transmission models. During the recent COVID-19 pandemic, many such data have been collected for the first time via newly developed DCM apps. However, it is unclear what the added value of these data is, unlike that of traditionally collected data via, e.g., surveys during non-epidemic times. The core idea behind our method is to express the number of infectious individuals as a function of the proportion of contacts that were with infected individuals and use this number as a starting point to initialize the remaining compartments of the model. As an important consequence, using our method, we can estimate key indicators such as the effective reproduction number using only two types of daily aggregated contact information, namely the average number of contacts and the average number of those contacts that were with an infected individual. We apply our method to the recent COVID-19 epidemic in the Netherlands, using self-reported data from the health surveillance app COVID RADAR and proximity-based data from the contact tracing app CoronaMelder. For both data sources, our corresponding estimates of the effective reproduction number agree both in time and magnitude with estimates based on other more detailed data sources such as daily numbers of cases and hospitalizations. This suggests that the use of DCM data in transmission models, regardless of the precise data type and for example via our method, offers a promising alternative for estimating the state of an epidemic, especially when more detailed data are not available.
\end{abstract}

\section{Introduction}

\paragraph{\bf Background.}
During the outbreak of an infectious disease, monitoring interactions between individuals provides useful information on the spread of the disease within the population. Information on social contacts has been used frequently in mathematical models of infectious disease transmission \cite{Beutels2006}. The two primary methods to collect this information have been diary-based social contact surveys \cite{Hoang2019} and field experiments involving proximity-measuring sensors (see \cite{Anglemyer20} and the references therein). Contact data obtained via either method have been successfully used to infer infection dynamics within a population and, in particular, to identify contact networks and heterogeneity in contact behavior \cite{Read2012}. However, most surveys and experiments are only conducted incidentally and not continuously on the daily level \cite{Verelst2021}. As a consequence, data from these sources is of limited use for continuously estimating short-term changes in key indicators during an ongoing epidemic.

Contact data from \emph{digital contact monitoring (DCM)} apps that were in use during an actual epidemic could potentially be used to address this limitation of current methods and sources of contact data. During the COVID-19 pandemic, many such apps have been developed and have continuously been collecting contact data. Here, we use the term DCM to refer to both digital contact tracing (DCT) and to public health surveillance via self-reporting of, e.g., symptoms and risk behavior. Most of these apps were developed to replace or complement manual contact tracing, in part because their use is less labor-intensive, more time-efficient, and highly scalable (see, e.g., \cite{Ferretti2020} and the references therein).

Despite the benefits of DCM compared to traditional methods and sources, it is currently unclear to what extent these contact data are useful in transmission modeling. For instance, there is no clear consensus on the effectiveness of DCT in terms of reducing the number of infections. On the one hand, multiple simulation studies find that the fraction of contacts captured by DCT increases approximately quadratically with the number of users \cite{Pozo2023, Barrat2021}. On the other hand, empirical evidence suggests that the proportion of asymptomatic cases among those detected purely as a consequence of DCT app use is small \cite{EU2022}. This implies that most infections detected via DCT could also have been detected in the absence of DCT, given a high willingness to test after symptom onset.

Another potential limitation in the use of contact data obtained from DCM apps is that the few contact data that are stored by DCT apps are very coarse and typically contain few or even no details on the registered contacts, such as user demographics. In particular, public versions of these DCT data are limited to counts on positive test results of DCT app users. This is because of the delicate balance between obtaining sufficient information on infectious contacts and preserving privacy of users \cite{Bradford2020}. For instance, individual location data from GPS-based DCT apps should be anonymized and aggregated so that individual travel patterns cannot be reconstructed \cite{Xu2017}. To comply with such requirements, most DCT apps that were developed during the COVID-19 epidemic had a decentralized design where as few privacy-sensitive data as possible were shared and stored \cite{Shahroz2021}.

\paragraph{\bf Our contributions.}
Altogether, the question remains to which extent DCM data is useful to inform transmission models. The goal of this paper is first of all to demonstrate the use of such data to initialize a conventional susceptible-exposed-infectious-recovered (SEIR) compartmental model. We then characterize the validity by estimating key indicators such as the effective reproduction number and the number of infectious individuals. We take the COVID-19 pandemic in the Netherlands as our use-case to compare our estimates with published values derived from detailed patient, testing and hospital data. 

Compared to other methods for estimating these quantities, our approach does not require model calibration or statistical estimation of transmission rates. Instead, we express the number of infectious individuals as a function of the proportion of contacts that were with infected individuals and initialize this value directly from the available data. Subsequently, we estimate the remaining compartments of the model and estimate the effective reproduction number using the method of \cite{Cori2013}. We also present an approximation for the effective reproduction number that only requires the data on contacts with infected individuals and the generation interval and thus is robust against any misspecifications in the latent, infectious, and immunity periods and the specific method used to initialize the compartmental model. This approximation is based on the assumption that the relative daily change in the number of susceptible individuals is small, which is particularly the case at the start of an epidemic when almost the entire population is susceptible.

We compare two different approaches to initialize the aggregated contact rate with infected individuals that use different methodologies for data collection. The first approach is based on self-reported data on close contacts and contacts with infected individuals that were collected via the public health surveillance app COVID RADAR \cite{vanDijk2021}, from April 2020 to February 2022. App users could, on a daily basis, submit a short questionnaire on COVID-19 related symptoms and risk behavior. In particular, users were asked to report whether they had been in contact with an infected individual in the preceding 14 days. For each reported contact with an infected individuals, we infer the exact date of contact by comparing consecutive answers to this question. 

In the second approach, we combine COVID-RADAR data with contact data obtained via the Dutch DCT app ``CoronaMelder'' \cite{TerHaar2023}, from October 2020 to April 2022. The primary goal of this app was to replace and later complement the manual contact tracing activities of the Dutch municipal health service. Due to the design of the app, the only publicly available contact data obtained via the app is the daily number of users that tested positive and notified their recent nearby contacts of this via the app. We combine these data with the COVID RADAR contact data to estimate the daily number of received notifications and, subsequently, use deconvolution to infer the expected time of these contacts.

We validate our approach and assumptions by comparing our estimates to those made by the Dutch National Institute for Public Health and the Environment (RIVM). These estimates were made continuously during the COVID-19 epidemic using detailed data on daily hospital admissions and on serological data from a population-based study. Finally, we conclude the paper by discussing the limitations of the approach and data, and providing recommendations for improving data collection with DCM apps like COVID-RADAR and CoronaMelder for the benefit of modeling infectious disease transmission in future epidemics.

\section{Methods}

\subsection{Data sources}
\label{sec_data_sources}

We use two data sources on contacts with infected individuals obtained from DCM apps:

\begin{description}
    \item[Self-reported contact data obtained via the COVID RADAR app \cite{vanDijk2020}] In summary, during the early phase of the pandemic (April 2020) the COVID RADAR app was launched after a short (social) media campaign. Users could anonymously report COVID-19 related symptoms, contacts and test results on a daily basis. In particular, users were asked to report the number of close interactions (within 1.5 meters) and whether they had a known recent contact with COVID-19 patients (in the last 14 days). These are the two variables that we will use in this research (see also Section~\ref{sec_data_int}). The question of the latter variable did not specify how close the contact must have been (e.g., within 1.5 meters). However, focus group interviews with app users did not indicate any feelings of ambiguity towards the phrasing of the question, unlike towards other contact-related questions \cite{Splinter2022}. Thus, we assume here that the reported recent contacts with COVID-19 patients were within 1.5 meters and thus could have contributed to transmission. To preserve privacy, each user was assigned a unique non-retraceable ID and data of that user was only linked to this ID. Moreover, the collection of precise personal information was minimized (e.g., instead of via date of birth, age was recorded in broader categories). Users were automatically reminded each other day to submit their responses and were offered insights into the data via in-app news updates and live feedback about personal (risk) behavior compared to the national average. In total over 250,000 unique individuals filled in the app over 8.5 million times, over a period of 750 days. For further details on the collection and validation of data via the COVID RADAR app, we refer to \cite{vanDijk2021}.

One major limitation of the data is that the user population is not a random sample of the Dutch population. In particular, the age distribution among users is not proportional to that of the Dutch population, with a disproportionately large share of middle-aged and elderly (60+) users. Not addressing this issue might bias the results since contact rates are known to differ between age groups and in particular are below average for people aged 60 or older \cite{Backer2021}. To account for this and improve the representativity of the user population for the entire Dutch population, user's responses were weighted in the rate calculations according to the proportion of their age group within the user population and within the Dutch population. Furthermore, to further reduce bias, users who listed ``healthcare professional'' as occupation were excluded. This is because people in these occupations are generally more in contact with COVID-19 patients but usually have a lower risk of infection due to the use of more elaborate protective equipment and a stricter adherence to measures.

\item[Contact tracing information obtained via the Dutch DCT app ``CoronaMelder'' \cite{MinVWS2022}] The CoronaMelder app was developed by the Dutch Ministry of Health, Welfare, and Sport and officially launched in October 2020 after several regional field tests in the summer of 2020. Users who downloaded and activated the app received a notification whenever another user that recently was nearby received a positive test result from an official test location of the Dutch municipal health service, i.e., whenever they were recently in close contact with a potentially infectious other user. The data contains the daily number of app downloads and the number of users who received a positive test and gave permission to send out notifications to other users. To improve users' privacy, no data was recorded on which users have received such a notification, meaning that no individual user data on risk-contacts is stored. As a consequence, and in contrast to the data from COVID RADAR, no other personal information is known on the user base of CoronaMelder in general and on those that gave permission in particular. Thus, unlike for the COVID RADAR data, we cannot correct for any bias in user characteristics such as age or occupation. The app was downloaded over 5,8 million times over a period of 606 days. We refer to \cite{TerHaar2023} for more technical details and an empirical evaluation of the effectiveness of the app.

In our approach, we require average numbers of notifications among active app users. However, daily estimates of this number have only been made from May 2021 onward and are not publicly available on the daily level \cite{Ebbers2021}. Furthermore, the number of app downloads is not representative of the number of active app users because many users stopped using the app throughout its run \cite{VanDerLaan2022}. To approximate the daily number of active users, we interpolate monthly published statistics on the number of active users \cite{CoronaMelder2022}. These reports indicate that, over time, the number of active users fluctuated between 1,9 and 3,4 million. To initialize the interpolation, we assume that the number of active users at the launch of CoronaMelder in October 2020 equals the cumulative number of app downloads at that time.
\end{description}

To demonstrate the value of the contact data from COVID RADAR and CoronaMelder, we will make comparisons with two other publicly available data sources that were used and published by RIVM during the COVID-19 epidemic \cite{RIVM}:
\begin{description}
    \item[Effective reproduction number.] 
The national-level effective reproduction number was estimated  based on state-reported numbers on hospital admissions as reported by the Dutch National Intensive Care Evaluation COVID-19 registry \cite{Dongelmans2022} (until June 12, 2020) and on the daily number of positive test results registered at the official test locations of the Dutch municipal health service (after June 12, 2020). For details on the methods used to estimate the reproduction number from these data, we refer to \cite{Wallinga2007,vandeKasteele2019}.

\item[Number of currently infectious individuals.]
The number of individuals that were infectious on a given day was estimated using age-stratified number of daily hospitalized COVID-19 patients in combination with serological data from a nationwide population based study \cite{Vos2021}. Here it is assumed that the infectious period ranges from two days before symptoms to four to eight days after symptoms. 
\end{description}

\subsection{Model}

In our approach, we model the spread of infection using the following standard deterministic SEIR-compartmental model:
\begin{align*}
\frac{d}{dt} S(t) &= -\frac{\beta(t) S(t) I(t)}{N} + \frac{R(t)}{\omega}, \\
\frac{d}{dt} E(t) &= \frac{\beta(t) S(t) I(t)}{N} - \frac{E(t)}{\alpha}, \\
\frac{d}{dt} I(t) &= \frac{E(t)}{\alpha} - \frac{I(t)}{\tau}, \\
\frac{d}{dt} R(t) &= \frac{I(t)}{\tau} - \frac{R(t)}{\omega}.
\end{align*}
Here, $S(t)$, $E(t)$, $I(t)$, and $R(t)$ denote the numbers of susceptible, exposed, infectious, and recovered individuals in the population at time $t$, respectively. Moreover, we denote by $ \omega$, $ \alpha$, and $\tau$ the average immunity, latent, and infectious period, respectively. We assume that the total size of the population is $N$ and remains constant through time. Finally, we decompose the transmission rate $\beta(t)$ as $\beta(t) = \varepsilon c(t)$, where $\varepsilon$ is the time-independent probability that a contact between a susceptible and an infectious individual leads to transmission and $c(t)$ is the contact rate at time $t$.

Our aim is to integrate data on contacts with infected individuals in this model. Therefore, it is crucial to explicitly focus in the SEIR model on the contacts with infectious individuals (in Section~\ref{sec_data_int}, we describe how we estimate contact rates with infectious individuals from contact rates with infected individuals). We denote this infectious contact rate by $c^I(t)$. The goal is to initialize the compartments of the SEIR model, given the infectious contact rate $c^I(t)$ and the overall contact rate $c(t)$. 

We focus on first initializing the number of infectious individuals $I(t)$. For this, we exploit the assumption of random mixing within the SEIR model, i.e., contacts between individuals happen at random and thus each pair of individuals is equally likely to meet. Under this assumption, the infectious and overall contact rate are related via $c^I(t) = c(t) \frac{I(t)}{N}$. This means that, given $c^I(t)$ and $c(t)$, we may initialize $I(t)$ as
\begin{equation}
I(t) = N \frac{c^I(t)}{c(t)}.
\label{eq_contact}
\end{equation}
This relationship implies that as long as the number of infectious individuals is nonzero, a positive fraction of the total number of contacts within the population is between a susceptible and an infectious individual. This fraction is represented by the fraction of contacts that were with an infectious person, which is modeled from the data as $c^I(t)/c(t)$. When this fraction is close to 1, almost all contacts were with infectious persons, meaning that within the SEIR model the share of infectious individuals in the total population of size $N$ is close to 1 as well. On the other hand, when the fraction is close to 0, hardly any contacts were with infectious persons, meaning that in the SEIR model the share of infectious individuals is close to 0 as well. We stress that this relation is invalided when there is a clear case of non-random mixing, for instance when measures of quarantine are in place. These are the moments where we expect our results to be less reliable and deviate most from other models and data sources that do take non-random mixing into account.

We use our model and the relation between the number of infectious individuals and the contact rates in (\ref{eq_contact}) to estimate the effective (instantaneous) reproduction number, where we apply the method of \cite{Cori2013}. This method requires information on the incidence, i.e., the number of new infections, and the generation interval, i.e., the time between the infection of an infected person and of their infector. The incidence at time $t$ is given in our model by $\mathcal{I}(t) = \beta(t)S(t)I(t)/N$ and the generation interval is assumed to be a random variable with a given probability density function $w$ (we discuss a suitable choice for the distribution of the generation interval in Section~\ref{sec_data_int}). We estimate the reproduction number at time $t$ as 
\[
\mathcal{R}(t) = \frac{\mathcal{I}(t)}{\int_{0}^{\infty} \mathcal{I}(t-i) w(i) di}.
\]
Using the relation in (\ref{eq_contact}), we may rewrite this expression for $\mathcal{R}(t)$ in terms of only the infectious contact rate, number of susceptible individuals, and the generation interval:
\begin{align}
    \mathcal{R}(t) &= \frac{\mathcal{I}(t)}{\int_{0}^{\infty} \mathcal{I}(t-i) w(i) di} \nonumber \\
&    =
    \frac{\frac{\beta(t) S(t) I(t)}{N}}
    {\int_{0}^{\infty} \frac{\beta(t-i) S(t-i) I(t-i)}{N} w(i) di} \nonumber \\
&    = \frac{c(t) S(t) I(t)}
    {\sum_{i=0}^{\infty} c(t-i) S(t-i) I(t-i) w(i)} \nonumber \\
    &    = \frac{c^I(t) S(t)}
    {\int_{0}^{\infty} c^I(t-i) S(t-i) w(i) di}. \label{eq_R}
\end{align}
To evaluate this expression, we require the time series of the size of the susceptible compartment up until time $t$. We compare two approaches for this, leading to two estimates $\mathcal{R}^1(t)$ and $\mathcal{R}^2(t)$ for the effective reproduction number:
\begin{enumerate}
    \item In the first approach, we initialize the entire SEIR model in the following way, based on the now known expression for $I(t)$ and direct manipulation of the differential equations that describe the model:
\begin{align*}
E(t) &= \alpha\left( \frac{d}{dt} I(t) + \frac{I(t)}{\tau} \right) , \\
R(t) & = e^{-\frac{t}{\omega}} R(0) + \int_{0}^{t} e^{-\frac{t-s}{\omega}} \frac{I(s)}{\tau} ds, \\
S(t) & = N - E(t) - I(t) - R(t).
\end{align*}
Substituting this expression in (\ref{eq_R}) yields our first estimate $\mathcal{R}^1(t)$ of the effective reproduction number.

\item For the second approach, we assume that the relative daily change in $S(t)$ is small, meaning that $S(t+i) \approx S(t)$ for small values of $i$. Substituting this relationship in (\ref{eq_R}) yields our second estimate $\mathcal{R}^2(t)$  of the effective reproduction number:
\begin{equation*}
    \mathcal{R}^2(t) =  \frac{c^I(t)}
    {\int_{0}^{\infty} c^I(t-i) w(i) di}.
\end{equation*}
In particular, this estimate of the effective reproduction number only depends on the infectious contact rate and on the generation interval.
\end{enumerate}

In general, note that neither $\mathcal{R}^1(t)$ nor $\mathcal{R}^2(t)$ depend on the unknown transmission probability $\varepsilon$. In Section~\ref{sec_result_R}, we compare both these approaches with each other and with the approach underlying the estimates produced by RIVM \cite{Wallinga2007,vandeKasteele2019}.

Finally, we note that, although we focus here on SEIR compartmental models, the core ideas and analyses presented in this section are also applicable to other types of compartmental models with a similar interaction between susceptible and infectious individuals.

\subsection{Data integration}
\label{sec_data_int}

Within the compartmental model, the values for the average latent, infectious, and immunity periods $\alpha$, $\tau$, and $\omega$ are chosen based on the literature as 5.5 days \cite{Xin2021}, 9.5 days \cite{Byrne2020}, and 90 days \cite{Stein2023}, respectively. The generation interval is assumed to follow a Gamma distribution with a mean of 4 days and a standard deviation of 2 days \cite{Backer2022}. To initialize the contact rate $c(t)$, we use the variable ``numberpersons150cm'' from the COVID RADAR dataset. This variable, here denoted by $C_i(t)$, states the number of persons that were within 150 cm of user $i$ on day $t$. We directly initialize $c(t)$ as the average of $C_i(t)$ over all $N$ users, i.e., $c(t) = \frac{1}{N} \sum_{i=1}^N C_i(t)$.

Finally, to initialize the infectious contact rate $c^I(t)$, we employ a two-step approach. First, we initialize an intermediate rate $z(t)$ of contacts with \emph{infected} individuals. In Sections~\ref{sec_int_radar} and~\ref{sec_int_melder}, we present two approaches for this using COVID RADAR data and CoronaMelder data, respectively. Second, we discuss how we transform this rate $z(t)$ of infected contacts to the desired rate $c^I(t)$ of infectious contacts. For this, note that we do not know the status of the infected persons at the time of the contact, i.e., whether they were actually infectious and could have caused a transmission. Therefore, we assume that the infected persons were either exposed or infectious and that this division is proportional to the relative difference between the latent and infectious period. This means that we assume a fraction $\frac{\alpha}{\alpha + \tau}$ of these contacts to be with an exposed individual and the remaining fraction $\frac{\tau}{\alpha + \tau}$ to be with an infectious individual. Thus, we initialize the infectious contact rate at time $t$ by appropriately discounting the rate $z(t)$ of contacts with infected individuals:
\begin{equation*}
    c^I(t) = \frac{\tau}{\alpha + \tau} z(t).
\end{equation*}

\subsubsection{Estimating the infected contact rate using COVID RADAR data}
\label{sec_int_radar}

We use the variable ``contact'' from the COVID RADAR dataset. This binary variable, denoted here by $C^{14}_i(t)$, represents whether or not user $i$ has had contact with an infected person within the 14 days before $t$. Note that the variables $C^{14}_i(t)$ individually do not provide precise information on when the contact with an infected person occurred. To extract this information from the data, we consider differences in reported values of $C^{14}_i(t)$ between subsequent days as follows. Suppose that a user $i$ reports on day $t$ that they had contact with an infected person within the last 14 days. If they report the next day $t+1$ that they did \emph{not} have such a contact within the last 14 days, then the contact that was reported on day $t$ must have occurred on day $t-14$ (assuming that this report concerns only one contact). Based on this reasoning, we construct a new intermediate variable $z_i(t)$ that is 1 when user $i$ had contact with an infected person on day $t$ and 0 otherwise:
\begin{equation*}
    z_i(t) = \begin{cases}
        1 & \text{if } C^{14}_i(t+14) = 1 \text{ and } C^{14}_i(t+15) = 0; \\
        0 & \text{otherwise.}
    \end{cases}
\end{equation*}
We assume that users report at most one contact per day with an infected person. This means that the average number of contacts with infected persons at day $t$ is given by $z(t) = \frac{1}{N} \sum_{i=1}^N z_i(t)$. 

Note that a reverse difference in subsequent reported values, i.e., when $C^{14}_i(t) = 0$ and $C^{14}_i(t+1) = 1$, does not necessarily mean that a contact with an infected person occurred at time $t-14$. This is because at the time of the contact, the user might not know yet that the other person was infected and only learn about this several days later.

 One limitation of the data is that users did not always submit a response daily. Therefore, we calculate the daily contact rates only over those users who submitted a response over that given day. Moreover, when a user has reported a recent infected contact at time $t$ but has not reported such a contact in any of the subsequent 14 days (either reporting not having had such a contact or not submitting a report at all), we assume that on day $t+1$ the user did not have an infected contact in the last 14 days. This means that we then initialize $z_i(t-14) = 1$.

\subsubsection{Estimating the infected contact rate using both COVID-RADAR and CoronaMelder data}
\label{sec_int_melder}

In this approach, we first estimate the daily number of people that have received a notification in the app and calculate from this the expected time series of total number of contacts with app users that recently tested positive. Unfortunately, no direct information is available on the number of received notifications due to the design of the app to preserve user privacy. Therefore, we instead use the variable ``Reported positive tests through app authorised by GGD (daily)'' from the CoronaMelder app data. This count variable, denoted here by $M(t)$, represents the number of app users who received a positive test result on day $t$ and agreed that other app users that were within 1.5m contact in the last 14 days receive a notification. To obtain a proxy for the daily number of users that receive a notification, we multiply $M(t)$ by the average number of contacts within 1.5m in the last 14 days according to the COVID RADAR data, i.e., with $\sum_{s=1}^{14} c(t-s)$.

To obtain an estimate of the daily number of infected contacts, we use deconvolution to delay the estimate of the daily number of received notifications by the time between contact with an infected individual and receiving a notification in the app. We model this time as a Gamma distribution with a mean and standard deviation of 3.1 and 3.6 days, respectively, which is in line with the distributions chosen in \cite{Leung2024} and the data analysis in \cite{Dolman2021} that both used a more detailed version of the CoronaMelder data.

\subsection{Evaluation setup}

We estimate the effective reproduction number and number of infectious individuals on the daily level, and not continuously, because all data sources we used are reported at this level of granularity, as described in Section~\ref{sec_data_sources}. This means that, when initializing the compartmental model, we use a time-discretized version of the SEIR model and of the probability density functions of the relevant distributions, i.e., those of the gneration interval and the time between contact with an infected individual and receiving a notification in the CoronaMelder app.

For the comparison of the effective reproduction number estimates, we select the time periods based on the range of the COVID RADAR and CoronaMelder data. This means that for the estimates purely based on COVID RADAR data, we consider the time period from April 2, 2020 to February 14, 2022 and for the estimates based partly on CoronaMelder data, we focus on the period from October 10, 2020 to February 14, 2022. The end dates of these periods are the same because the data collection for the COVID RADAR app ended on February 28, 2022. For the comparison of the estimates of the number of infectious individuals, we use the same time periods as for the estimates of the effective reproduction number but note that the estimates from RIVM are available only until June 28, 2021. Within the selected time periods, we focus in particular on a cluster of national-wide super-spreading events at the start of July 2021 that became known as the ``dancing with Janssen'' event \cite{vanderVeer2023}. This event cluster occurred after a change in policy that allowed people to directly visit bars and nightclubs without any distancing restrictions after receiving one dose of the Janssen vaccination.

We first compare the two different expressions $\mathcal{R}^1(t)$ and $\mathcal{R}^2(t)$ to estimate the effective reproduction number. More precisely, we validate the assumption that $S(t + i) \approx S(t)$ for small $i$ and thus that the simpler expression $\mathcal{R}^2(t)$ is to be preferred. In a second step, we compare our estimates of the effective reproduction number and the number of infectious individuals with those made by RIVM. To reduce the impact of noisy data, we smooth the infectious and overall contact rates by fitting for each rate a cubic smoothing spline function, where we select the smoothing parameter via generalized cross validation \cite{Wahba1990}.

All source code underlying the evaluation is written in Python version 3.9.13 and is publicly available at \url{https://github.com/mhhschootuiterkamp/DCM-data-integration-in-SEIR-models}.

\section{Evaluation}
\label{sec_result_R}

\subsection{Results}

\begin{figure}[t!]
    \centering
    \includegraphics[scale = .6]{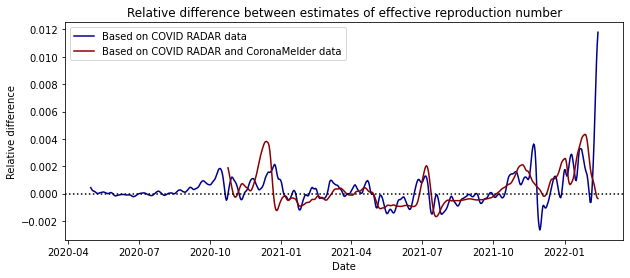}
    \caption{Relative difference between our estimates of the effective reproduction number ($\mathcal{R}^2(t) 
    / \mathcal{R}^1(t) - 1$).}
    \label{fig:R_rel}
\end{figure}

Figure~\ref{fig:R_rel} shows the relative difference between our two estimates $\mathcal{R}^1(t)$ and $\mathcal{R}^2(t)$ of the effective reproduction number, i.e., $\mathcal{R}^2(t) / \mathcal{R}^1(t) - 1$. Regardless of the used data source (COVID RADAR or CoronaMelder), this difference is generally below 0.4\%, with a single outlier at the very last day for the estimate based on only COVID RADAR. This suggests that the estimates $\mathcal{R}^1(t)$ and $\mathcal{R}^2(t)$ are very close together and thus that the assumption used to derive $\mathcal{R}^2(t)$ is justified. Therefore, in the remainder of this section, we only use the approximation $\mathcal{R}^2(t)$ when comparing our estimates to those of RIVM because this estimate is robust against misspecifications in the overall contact rate $c(t)$ and the average immunity, latent, and infectious period and the specific method used to initialize the compartmental model.

\begin{figure}[t!]
    \centering
    \includegraphics[scale = .6]{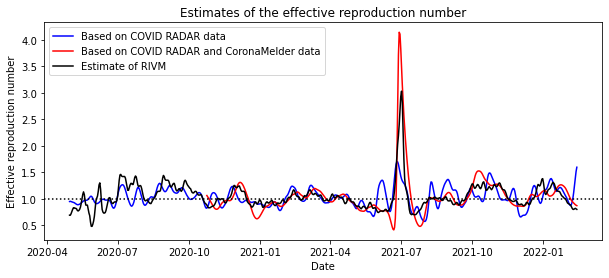}
    \caption{Estimates of the effective reproduction number.}
    \label{fig:R}
\end{figure}

\begin{figure}[t!]
    \centering
    \includegraphics[scale = .6]{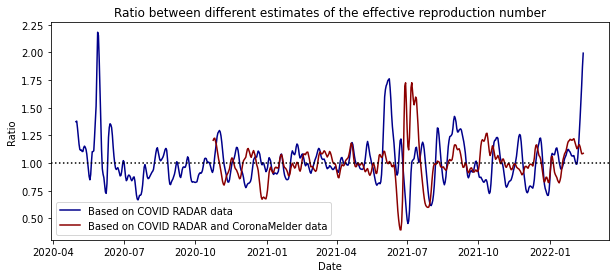}
    \caption{Ratios between our effective reproduction number estimates and the estimate of RIVM.}
    \label{fig:R_r}
\end{figure}

Figure~\ref{fig:R} shows the effective reproduction numbers as estimated via our method ($\mathcal{R}^2(t)$, using either COVID RADAR or CoronaMelder contact data) and as estimated by RIVM (using hospitalization and positive test data). Overall, all three estimates show the same general trend and, as shown in Figure~\ref{fig:R_r}, our estimates generally deviate from those of RIVM by at most 25\%. In particular, all three estimates reflect the strong increase in new cases around the start of July 2021 as a consequence of the ``dancing with Janssen'' cluster of super-spreading events mentioned before. However, Figures~\ref{fig:R} and~\ref{fig:R_r} show that the three estimates disagree on the distribution of infections over time during these events. The approach based on COVID RADAR estimates these infections to occur in two peaks, of which one occurs earlier than the single peak estimate by RIVM. On the other hand, the approach based on both COVID RADAR and CoronaMelder data estimates the single peak in infections to occur later than the RIVM estimate.

\begin{figure}[t!]
    \centering
    \includegraphics[scale = .6]{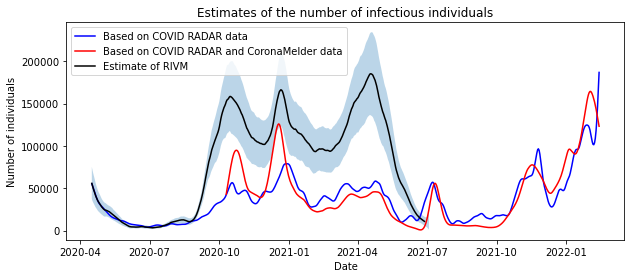}
    \caption{Estimates of the number of infectious individuals, including the 95\% confidence interval for the estimate of RIVM.}
    \label{fig:prevalence}
\end{figure}

\begin{figure}[t!]
    \centering
    \includegraphics[scale = .6]{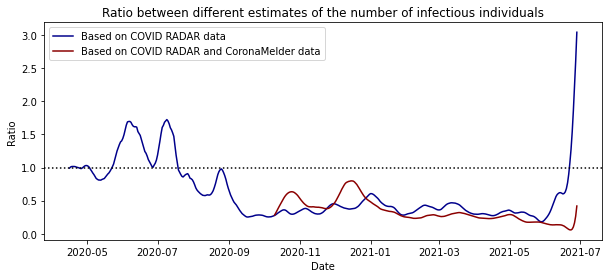}
    \caption{Ratios between our estimates of the number of infectious individuals and the estimate of RIVM.}
    \label{fig:prevalence_r}
\end{figure}

Figure~\ref{fig:prevalence} shows the number of infectious individuals as estimated by our methods and by RIVM, and Figure~\ref{fig:prevalence_r} shows the ratios between our estimates and those of RIVM. The results in Figure~\ref{fig:prevalence} indicate that all three estimates display the same general patterns and waves of infections. However, Figure~\ref{fig:prevalence_r} shows a distinct behavior between different time periods in the magnitude of the estimates. From April to September 2020, the estimate based on COVID RADAR data is in the same order of magnitude as the RIVM estimate and often larger.  However, from September 2020 on, both estimates increase and from October 2020 on, their proportion remains stable at 0.41 on average. The ratio between the RIVM estimate and that based on both COVID RADAR and CoronaMelder data fluctuates a bit more and is on average 0.34. The sudden increase in this ratio a at the end of June 2021 is due to the apparently early identification of the ``dancing with Janssen'' cluster of super-spreading events in July 2021.

\subsection{Discussion of the results}

The observation in Figure~\ref{fig:R} that all three estimates of the effective reproduction number follow the same trend underscores the added value of models that do not require fitting transmission rates to observed output. More precisely, population-level insight in the status of the epidemic that is obtained directly from the available population-aggregated self-reported and proximity-based contact data is similar to insight obtained using more detailed data and statistically estimated transmission parameters. We did observe, however, that smoothing was indeed necessary to obtain interpretable results, i.e., the contact data is sparse and noisy.

 Regarding the effective reproduction number, both our estimates and those of RIVM detected the increase in infections as a consequence of the super-spreading events in July 2021, but disagree on the precise magnitude and distribution of infections over time. One reason for the relatively smaller peak estimates based on COVID RADAR data is that the rise in new cases during this super-spreading event was much stronger among younger age groups. Because these groups are strongly underrepresented in the COVID RADAR user base, the users within these groups are likely to be unrepresentative of the entire age group \cite{vanDijk2021} and the reweighing does not account for this type of bias.

Regarding the number of infectious individuals, both our estimates generally agree on both the timing and magnitude. Given that in both estimates the average number of contacts in (\ref{eq_contact}) is estimated using purely COVID RADAR data, also the estimates of the infected contact rates generally agree. This suggests that COVID RADAR and CoronaMelder detect a similar volume of infected contacts. One exception where the estimates differ is the two peaks between October and December 2020. One reason for this could be the method for approximating the number of active users of CoronaMelder during this period. As mentioned in Section~\ref{sec_data_sources}, only from May 2021 onward estimates of the number of active users have been made. Between October 2020 and May 2021, our approximation is a linear interpolation between the number of downloads at October 10, 2020 and the first publicly reported number of active users on April 30, 2021. The strong deviation of our estimates in the period Ocotber - December 2020 could thus indicate a disproportional change in active users that is not captured by the interpolation.

From October 2020 on our estimates of the number of infectious individuals are much lower than those of RIVM. One possible explanation for this is that the RIVM estimates are obtained by extrapolating the relations between age-stratified serological data and hospitalization in the period before October 2020.  However, these relations could alter during the course of the pandemic due to, e.g., vaccination and hospital crowding and available capacity. In particular, new SARS-CoV-2 variants may have changed these relations due to a change in the share of and heterogeneity among COVID-19 patients who require hospitalization \cite{deBoer2023} and increased transmissibility \cite{Campbell2021}. Another potential reason is that the RIVM estimate is less biased against asymptomatic cases. These cases are underrepresented in positive test results and thus contacts with asymptomatic infected persons are less likely to be present in both the COVID RADAR data and CoronaMelder data. In particular, the ratios in Figure~\ref{fig:prevalence_r} are of comparable size to the estimated share of symptomatic COVID-19 cases in the Netherlands \cite{McDonald2021}.

Finally, we note that, generally, a strong violation of the random mixing assumption within the compartmental model may cause our estimate to deviate from the true number of infectious individuals. Recall that we initialize the number of infectious individuals as the share of contacts that are with infectious individuals, scaled up by the population size (\ref{eq_contact}). This share can change when at the same time the total number of infectious individuals remains the same, for instance when there is a change in the proportion of infectious individuals that quarantine themselves.

\section{Discussion on methods}

Below we discuss limitations regarding the accuracy and user group bias in both the self-reported contact data from COVID RADAR and the DCT data from CoronaMelder and regarding the modeling of transmission that may have affected the results.

\paragraph{\bf Data accuracy.}
The data obtained via the COVID RADAR app is self-reported, meaning that the accuracy of the data depends on user's awareness and recollection of contacts and on their truthfulness and precision in reporting these contacts. Asymptomatic cases of COVID-19 often remain untested and users are therefore less likely to be aware of contacts with infected but asymptomatic individuals \cite{McDonald2021}. Moreover, short contacts with, e.g., infected strangers are less likely to be reported because these encounters are more likely to be forgotten \cite{Smieszek2012}. Regarding truthfulness, focus group interviews with participants of the app indicate that users were generally motivated to participate out of altruism, collectivism, and a desire to help science in developing useful public health surveillance tools  \cite{Splinter2022}. Thus, it seems unlikely that users purposefully submit untruthful responses on a large scale. However, regarding precision, these interviews also suggest that the formulation of the questions could be simplified or made more precise. In particular, users may interpret questions differently, which reduces the reliability of the given answers. The CoronaMelder app, like many other DCT apps, used BlueTooth Low Energy to detect other nearby devices that had the app installed \cite{TerHaar2023}. However, the proximity detection range for this type of BlueTooth is not fixed and may range from 2 to 10 meters, where the strength of the signal depends on device orientation and obstacles such as indoor walls \cite{Leith2020}. This results in an unknown number of both false positives and negatives that influences the number of app users that receive a notification.

\paragraph{\bf User group bias.}
For both COVID RADAR and CoronaMelder, app users were not selected via a random sampling procedure. This complicates statistical analysis of the questionnaire results and DCT notification data. To reduce potential bias in the results, we applied post-stratification with regard to age to the COVID RADAR data in a pre-processing step. However, a more sophisticated sampling and stratification procedure would be required to quantify the uncertainty in our estimates, for instance via prediction or confidence intervals. Furthermore, bias in both DCM data sources and those used by RIVM may also occur because of factors that are known to influence a person's willingness to get tested for COVID-19 or to use DCM apps. Some of these factors impact both aspects, such as nationality or socioeconomic status (see \cite{McDonald2021b, Ritsema2022} and the references therein), but others are more complementary. For instance, on the one hand, the level of literacy \cite{Splinter2022} and attitude towards technology in general \cite{Jansen2021} specifically impact the uptake of DCM apps. On the other hand, the availability of testing locations near one's home and the ability to visit test locations located further away via, e.g., public transport specifically impact the willingness to test (see \cite{Vink2022} and the references therein). The observation that the different estimates of the reproduction number generally agree suggests that the methods underlying these estimates are robust against such complementary biases in the input data.

\paragraph{\bf Model.}
Finally, for the modeling of infection transmission, we used a standard SEIR compartmental model that assumes random mixing and does not include any population heterogeneity or underlying contact network structure. This is a strong simplification of existing heterogeneity in contacts \cite{Mossung2008} and of the over-dispersed nature of COVID-19 transmission in general and of super-spreading events in particular \cite{Sneppen2021}. However, our goal was to show the added value of using coarse contact data that lacks this heterogeneity, which motivated our choice for a transmission model that does not require this. Moreover, our population-level estimates of the effective reproduction number generally agree with those of RIVM, which are based on more complex age-stratified compartmental models. Thus, our results suggest that simple models might be sufficient to monitor the population-level progression of an epidemic, despite the biases and lack of heterogeneity in the used contact data. We do believe that including overdispersion in the modeling of population-level contact behavior might improve the reliability of our estimates during super-spreading events and leave this as a direction for future research.

\section{Conclusions and recommendations}

The goal of this paper was to demonstrate how aggregated data on contacts with infected individuals, obtained from digital contact monitoring (DCM) apps, can inform mathematical transmission models and can be used to estimate key indicators such as the effective reproduction number. For this, we integrated both self-reported contact data from COVID RADAR \cite{vanDijk2021} and contact tracing information from the Dutch digital contact tracing (DCT) app CoronaMelder \cite{MinVWS2022} in a standard susceptible-exposed-infectious-recovered compartmental model. We showed that, despite any biases in the underlying contact data, the estimates that our approach produces agree with those made using more detailed data on hospitalizations and seroprevalence research. 

We conclude this paper with several recommendations for improvements in DCM apps such as COVID RADAR and CoronaMelder that may increase their value in monitoring and modeling the state of an epidemic. Our approach requires only two daily aggregated types of contact information, namely the average number of overall contacts that could facilitate transmission for the infectious disease in question and the average number of such contacts that were with an infected individual. The success of our approach depends on the availability and quality of these two quantities. Thus, our recommendation is to include the collection of this information in future apps.

 The COVID RADAR app already provides information on overall contacts directly. However, information on contacts with infected individuals had to be inferred by comparing consecutive reports from the same individual on whether such a contact occurred within the last 14 days. This means that only after 14 days we can infer the exact time of a contact with an infected person. As a consequence, we can only estimate the effective reproduction number with a delay of 14 days, which limits the usability of the method during an ongoing epidemic where there is a high urgency to obtain timely estimates. To obtain more timely data on the time of infected contacts and thus allow for more timely estimates of the effective reproduction number, we therefore recommend that a future version of the app includes an additional question that asks for the specific date on which the contact with the infected individual occurred.

Regarding the DCT app CoronaMelder, no data was available on the daily number of close contacts that a given user had with other users who installed the app. However, this information was actually recorded and temporarily stored on individual users' devices. In fact, every day, each device compared these contacts with a daily distributed list of positively tested app users and notified their user if there was a match. In the current app design, the ``local'' lists of contacts stayed on the individual device and were deleted after 14 days. However, we recommend that in a next version of the app the number of contacts is communicated back and aggregated on a daily basis to provide an estimate for the overall average contact rate.  Moreover, also the daily number of matches should be communicated and the date on which the corresponding contact occurred. Using these aggregated data would improve the estimation of the daily number of received notifications, which we now resolved by incorporating data from COVID RADAR. The current design of the app did not record and store this information due to its decentralized design. However, similarly to the COVID RADAR app, a privacy-by-design approach could be adapted to keep track of this information without violating individual users' privacy. In particular, no information on individual users other than their number of contacts would need to be shared, which is already less information than what is shared in the COVID RADAR app.

\section*{Acknowledgments}
This work is supported by Netherlands Organisation for Scientific Research (NWO) through ZonMw grants no. 10430032010011, 10430372310021, and 10710062310008. R.v.d.H. and N.L. are also supported by NWO through Gravitation NETWORKS grant no. 024.002.003.

\bibliographystyle{vancouver}
\bibliography{lib_DCM}

\begin{thebibliography}{10}

\bibitem{Beutels2006}
Beutels P, Shkedy Z, Aerts M, van Damme P.
\newblock Social mixing patterns for transmission models of close contact
  infections: exploring self-evaluation and diary-based data collection through
  a web-based interface.
\newblock Epidemiol Infect. 2006;134(6):1158–1166.

\bibitem{Hoang2019}
Hoang T, Coletti P, Melegaro A, Wallinga J, Grijalva CG, Edmunds JW, et~al.
\newblock A systematic review of social contact surveys to inform transmission
  models of close-contact infections.
\newblock Epidemiol. 2019;30(5):723--736.

\bibitem{Anglemyer20}
Anglemyer A, Moore T, Parker L, Chambers T, Grady A, Chiu K, et~al.
\newblock Digital contact tracing technologies in epidemics: a rapid review.
\newblock Cochrane Database Syst Rev. 2020;(8).

\bibitem{Read2012}
Read JM, Edmunds WJ, Riley S, Lessler J, Cummings DAT.
\newblock Close encounters of the infectious kind: methods to measure social
  mixing behaviour.
\newblock Epidemiol Infect. 2012;140(12):2117–2130.

\bibitem{Verelst2021}
Verelst F, Hermans L, Vercruysse S, Gimma A, Coletti P, Backer JA, et~al.
\newblock {SOCRATES-CoMix}: A platform for timely and open-source contact
  mixing data during and in between COVID-19 surges and interventions in over
  20 European countries.
\newblock BMC Med. 2021;19(254).

\bibitem{Ferretti2020}
Ferretti L, Wymant C, Kendall M, Zhao L, Nurtay A, Abeler-Dörner L, et~al.
\newblock Quantifying SARS-CoV-2 transmission suggests epidemic control with
  digital contact tracing.
\newblock Sci. 2020;368(6491):eabb6936.

\bibitem{Pozo2023}
Pozo-Martin F, {Beltran Sanchez} MA, Müller SA, Diaconu V, Weil K, {El
  Bcheraoui} C.
\newblock Comparative effectiveness of contact tracing interventions in the
  context of the COVID-19 pandemic: a systematic review.
\newblock Eur J Epidemiol. 2023;38:243--266.

\bibitem{Barrat2021}
Barrat A, Cattuto C, Kivelä M, Lehmann S, Saramäki J.
\newblock Effect of manual and digital contact tracing on COVID-19 outbreaks: a
  study on empirical contact data.
\newblock J R Soc Interface. 2021;18(178):20201000.

\bibitem{EU2022}
Prodan A, Birov S, von Wyl V, , Ebbers W.
\newblock Digital contact tracing study: Study on lessons learned, best
  practices and epidemiological impact of the common European approach on
  digital contact tracing to combat and exit the COVID-19 pandemic.
\newblock European Commission; 2022.
\newblock Available from:
  \url{https://digital-strategy.ec.europa.eu/en/library/covid-19-digital-contact-tracing-study}.

\bibitem{Bradford2020}
Bradford L, Aboy M, Liddell K.
\newblock {COVID-19 contact tracing apps: a stress test for privacy, the GDPR,
  and data protection regimes}.
\newblock J Law Biosci. 2020 05;7(1):lsaa034.

\bibitem{Xu2017}
Xu F, Tu Z, Li Y, Zhang P, Fu X, Jin D.
\newblock Trajectory recovery from ash: User privacy Is NOT preserved in
  aggregated mobility data.
\newblock In: Proceedings of the 26th International Conference on World Wide
  Web. WWW '17. Republic and Canton of Geneva, CHE: International World Wide
  Web Conferences Steering Committee; 2017. p. 1241–1250.

\bibitem{Shahroz2021}
Shahroz M, Ahmad F, Younis MS, Ahmad N, {Kamel Boulos} MN, Vinuesa R, et~al.
\newblock COVID-19 digital contact tracing applications and techniques: A
  review post initial deployments.
\newblock Transp Eng. 2021;5:100072.

\bibitem{Cori2013}
Cori A, Ferguson NM, Fraser C, Cauchemez S.
\newblock A new framework and software to estimate time-varying reproduction
  numbers during epidemics.
\newblock Am J Epidemiol. 2013 09;178(9):1505--1512.

\bibitem{vanDijk2021}
van Dijk WJ, Saadah NH, Numans ME, Aardoom JJ, Bonten TN, Brandjes M, et~al.
\newblock COVID RADAR app: Description and validation of population
  surveillance of symptoms and behavior in relation to COVID-19.
\newblock PLOS ONE. 2021 06;16(6).

\bibitem{TerHaar2023}
Ter~Haar W, Bosdriesz J, Venekamp RP, Schuit E, van~den Hof S, Ebbers W, et~al.
\newblock The epidemiological impact of digital and manual contact tracing on
  the SARS-CoV-2 epidemic in the Netherlands: Empirical evidence.
\newblock PLOS Digit Health. 2023 12;2(12):1--17.

\bibitem{vanDijk2020}
{van Dijk} WJ. COVID RADAR app.
\newblock DANS; 2020.
\newblock Dataset available from \url{https://doi.org/10.17026/dans-zcd-m9dh}.

\bibitem{Splinter2022}
Splinter B, Saadah NH, Chavannes NH, Kiefte-de Jong JC, Aardoom JJ.
\newblock Optimizing the acceptability, adherence, and inclusiveness of the
  COVID Radar surveillance app: Qualitative study using focus groups, thematic
  content analysis, and usability testing.
\newblock JMIR Form Res. 2022;6(9):e36003.

\bibitem{Backer2021}
Backer JA, Mollema L, Vos ERA, Klinkenberg D, van~der Klis FRM, de~Melker HE,
  et~al.
\newblock Impact of physical distancing measures against COVID-19 on contacts
  and mixing patterns: repeated cross-sectional surveys, the Netherlands,
  2016–17, April 2020 and June 2020.
\newblock Eurosurveill. 2021;26(8):2000994.

\bibitem{MinVWS2022}
{Ministerie van Volksgezondheid, Welzijn en Sport}. CoronaMelder statistics;.
\newblock Dataset available from
  \url{https://github.com/minvws/nl-covid19-notification-app-statistics}.

\bibitem{Ebbers2021}
Ebbers W, Hooft L, van~der Laan N, Metting E.
\newblock Evaluatie CoronaMelder: Een overzicht na 9 maanden.
\newblock Ministerie van Volksgezondheid, Welzijn en Sport; 2021.
\newblock Accessed on January 11, 2025.
\newblock Available from:
  \url{https://www.rijksoverheid.nl/documenten/publicaties/2021/05/28/rapporten-evaluatie-coronamelder-9-maanden}.

\bibitem{VanDerLaan2022}
van~der Laan LN, de~Wit JMS.
\newblock Factoren relevant voor het stoppen met gebruik van de CoronaMelder.
\newblock Tilburg University; 2022.
\newblock Available from:
  \url{https://research.tilburguniversity.edu/en/publications/factoren-relevant-voor-het-stoppen-met-gebruik-van-de-coronamelde}.

\bibitem{CoronaMelder2022}
{Ministerie van Volksgezondheid, Welzijn en Sport}. CoronaMelder data
  dashboard; 2022.
\newblock Accessed on January 11, 2025 via Archive.today.
\newblock Available from:
  \url{https://archive.ph/2022.04.06-092154/https:/coronamelder.nl/nl/faq/1-13-coronamelder-data-dashboard/}.

\bibitem{RIVM}
{Rijksinstituut voor Volksgezondheid en Milieu}. COVID-19 dataset;.
\newblock Dataset available from \url{https://data.rivm.nl/covid-19/}.

\bibitem{Dongelmans2022}
Dongelmans DA, Termorshuizen F, Brinkman S, Bakhshi‑Raiez F, Arbous MS,
  de~Lange DW, et~al.
\newblock Characteristics and outcome of COVID-19 patients admitted to the ICU:
  A nationwide cohort study on the comparison between the first and the
  consecutive upsurges of the second wave of the COVID-19 pandemic in the
  Netherlands.
\newblock Ann Intensive Care. 2022;12(5).

\bibitem{Wallinga2007}
Wallinga J, Lipsitch M.
\newblock How generation intervals shape the relationship between growth rates
  and reproductive numbers.
\newblock Proc R Soc B Biol Sci. 2007;274(1609):599--604.

\bibitem{vandeKasteele2019}
van~de Kassteele J, Eilers PHC, Wallinga J.
\newblock Nowcasting the number of new symptomatic cases during infectious
  disease outbreaks using constrained p-spline smoothing.
\newblock Epidemiol. 2019;30:737--745.

\bibitem{Vos2021}
Vos ERA, den Hartog G, Schepp RM, Kaaijk P, van Vliet J, Helm K, et~al.
\newblock Nationwide seroprevalence of {SARS-CoV-2} and identification of risk
  factors in the general population of the Netherlands during the first
  epidemic wave.
\newblock J Epidemiol Community Health. 2021;75:489--495.

\bibitem{Xin2021}
Xin H, Li Y, Wu P, Li Z, Lau EHY, Qin Y, et~al.
\newblock {Estimating the latent period of coronavirus disease 2019
  (COVID-19)}.
\newblock Clin Infect Dis. 2021 09;74(9):1678--1681.

\bibitem{Byrne2020}
Byrne AW, McEvoy D, Collins AB, Hunt K, Casey M, Barber A, et~al.
\newblock Inferred duration of infectious period of SARS-CoV-2: Rapid scoping
  review and analysis of available evidence for asymptomatic and symptomatic
  COVID-19 cases.
\newblock BMJ Open. 2020;10(8).

\bibitem{Stein2023}
Stein C, Nassereldine H, Sorensen RJD, Amlag JO, Bisignano C, Byrne S, et~al.
\newblock Past SARS-CoV-2 infection protection against re-infection: a
  systematic review and meta-analysis.
\newblock Lancet. 2023;401(10379):833--842.

\bibitem{Backer2022}
Backer JA, Eggink D, Andeweg SP, Veldhuijzen IK, van Maarseveen N, Vermaas K,
  et~al.
\newblock Shorter serial intervals in SARS-CoV-2 cases with Omicron BA.1
  variant compared with Delta variant, the Netherlands, 13 to 26 December 2021.
\newblock Eurosurveill. 2022;27(6).

\bibitem{Leung2024}
Leung KY, Metting E, Ebbers W, Veldhuijzen I, Andeweg SP, Luijben G, et~al.
\newblock Effectiveness of a COVID-19 contact tracing app in a simulation model
  with indirect and informal contact tracing.
\newblock Epidemics. 2024;46:100735.

\bibitem{Dolman2021}
Dolman TL.
\newblock Tabellenrapport CoronaMelder GGD GHOR Nederland.
\newblock GGD GHOR Nederland; 2021.
\newblock Accessed on October 16, 2024.
\newblock Available from:
  \url{https://www.rijksoverheid.nl/documenten/publicaties/2021/05/28/rapporten-evaluatie-coronamelder-9-maanden}.

\bibitem{vanderVeer2023}
van~der Veer BMJW, Gorgels KMF, den Heijer CDJ, Hackert V, van Alphen LB,
  Savelkoul PHM, et~al.
\newblock SARS-CoV-2 transmission dynamics in bars, restaurants, and
  nightclubs.
\newblock Front Microbiol. 2023;14.

\bibitem{Wahba1990}
Wahba G.
\newblock Estimating the Smoothing Parameter.
\newblock In: Spline Models for Observational Data. Philadelphia, PA: Society
  for Industrial and Applied Mathematics; 1990. p. 45--65.

\bibitem{deBoer2023}
de~Boer PT, van~de Kassteele J, Vos ERA, van Asten L, Dongelmans DA, van
  Gageldonk-Lafeber AB, et~al.
\newblock Age-specific severity of severe acute respiratory syndrome
  coronavirus 2 in February 2020 to June 2021 in the Netherlands.
\newblock Influenza Other Respir Viruses. 2023;17(8):e13174.

\bibitem{Campbell2021}
Campbell F, Archer B, Laurenson-Schafer H, Jinnai Y, Konings F, Batra N, et~al.
\newblock Increased transmissibility and global spread of SARS-CoV-2 variants
  of concern as at June 2021.
\newblock Eurosurveill. 2021;26(24).

\bibitem{McDonald2021}
McDonald SA, Miura F, Vos ERA, van Boven M, de~Melker HE, van~der Klis FRM,
  et~al.
\newblock Estimating the asymptomatic proportion of SARS-CoV-2 infection in the
  general population: Analysis of nationwide serosurvey data in the
  Netherlands.
\newblock Eur J Epidemiol. 2021;36:735--739.

\bibitem{Smieszek2012}
Smieszek T, Burri EU, Scherzinger R, Scholz RW.
\newblock Collecting close-contact social mixing data with contact diaries:
  reporting errors and biases.
\newblock Epidemiol Infect. 2012;140(4):744--752.

\bibitem{Leith2020}
Leith DJ, Farrell S.
\newblock Coronavirus contact tracing: evaluating the potential of using
  bluetooth received signal strength for proximity detection.
\newblock SIGCOMM Comput Commun Rev. 2020;50(4):66–74.

\bibitem{McDonald2021b}
McDonald SA, Soetens LC, Schipper CMA, Friesema I, {van den Wijngaard} CC,
  Teirlinck A, et~al.
\newblock Testing behaviour and positivity for {SARS-CoV-2} infection: insights
  from webbased participatory surveillance in the Netherlands.
\newblock BMJ Open. 2021;11:e056077.

\bibitem{Ritsema2022}
Ritsema F, Bosdriesz JR, Leenstra T, Petrignani MWF, Coyer L, Schreijer AJM,
  et~al.
\newblock Factors associated With using the COVID-19 mobile contact-tracing app
  among individuals diagnosed with SARS-CoV-2 in Amsterdam, the Netherlands:
  Observational study.
\newblock JMIR Mhealth Uhealth. 2022;10(8):e31099.

\bibitem{Jansen2021}
Jansen-Kosterink S, Hurmuz M, den Ouden M, van Velsen L.
\newblock Predictors to use mobile apps for monitoring COVID-19 symptoms and
  contact tracing: Survey among Dutch citizens.
\newblock JMIR Form Res. 2021;5(12):e28416.

\bibitem{Vink2022}
Vink M, Igl\'{o}i Z, Fanoy EB, {van Beek} J, Boelsums T, {de Graaf} M, et~al.
\newblock Community-based SARS-CoV-2 testing in low-income neighbourhoods in
  Rotterdam: Results from a pilot study.
\newblock J Glob Health. 2022;1(12:05042).

\bibitem{Mossung2008}
Mossong J, Hens N, Jit M, Beutels P, Auranen K, Mikolajczyk R, et~al.
\newblock Social contacts and mixing patterns relevant to the spread of
  infectious diseases.
\newblock PLOS Med. 2008 03;5(3):e74.

\bibitem{Sneppen2021}
Sneppen K, Nielsen BF, Taylor RJ, Simonsen L.
\newblock Overdispersion in COVID-19 increases the effectiveness of limiting
  nonrepetitive contacts for transmission control.
\newblock Proc Nat Acad Sci. 2021;118(14):e2016623118.

\end{thebibliography}

\end{document}